\documentclass{article}%
\usepackage{amsmath}
\usepackage{amsfonts}
\usepackage{amssymb}
\usepackage{graphicx}%
\providecommand{\U}[1]{\protect\rule{.1in}{.1in}}

\begin{document}

\title{Comment on \textquotedblleft On Phase Selective Quantum
Eraser\textquotedblright\ (arXiv:1501.00817 [quant-ph])}
\author{Luiz Carlos Ryff\\\textit{Instituto de F\'{\i}sica, Universidade Federal do Rio de Janeiro,}\\\textit{Caixa Postal 68528, 21041-972 Rio de Janeiro, Brazil}\\E-mail: ryff@if.ufrj.br}
\maketitle

\begin{abstract}
In a recent interesting article A. Heuer, G. Pieplow, and R. Menzel discuss a
quantum-eraser experiment to investigate the complementarity of wave-like and
particle-like behavior of photons. I would like to draw your attention to the
fact that the very same experimental setup was suggested in a previous paper,
and take advantage of this opportunity to examine some aspects of this
controversial subject.

\end{abstract}

In a recent interesting article \textrm{[1]} A. Heuer, G. Pieplow, and R.
Menzel (HPM) discuss a quantum-eraser experiment. To \textquotedblleft
demonstrate a quantum eraser protocol for the complementarity of wave-like and
particle-like behavior of photons\textquotedblright\ they introduce an extra
Mach-Zehnder interferometer in an experiment on induced coherence without
induced emission \textrm{[2]}. In fact, the very same experimental setup was
suggested in a previous paper \textrm{[3]. }In the HPM experiment, whenever it
is possible to know which crystal emits the pair of photons, there is no
interference. On the other hand, when this information is erased, interference
can be observed.\textrm{\ }Since the concept of the quantum eraser was
introduced \textrm{[4]}, many experiments have been proposed and performed
\textrm{[5]}. What makes the experiment suggested in \textrm{[3]} particularly
interesting is that we have a situation in which, apparently, the idler and
signal photons are emitted by different crystals.\textrm{\ }Since the pair of
photons is supposed to be generated when a photon of the pumping laser is
split in two inside a crystal (an idler and a signal), we are facing a
mind-boggling situation. This strongly indicates that, as emphasized by Lamb
\textrm{[6]}, we have to be very careful when dealing with the photon
concept.\textrm{\ }That a single photon can be spread throughout the
electromagnetic field of independent sources was already implicit in an
experiment on the interference of attenuated light emitted by independent
lasers \textrm{[7]}, even though it involved no authentic one photon (Fock)
states, as would be the ideal. It is well established that a photon can
disclose wave-like and particle-like features. On the other hand, the idea
that a photon can \textit{behave} as a wave \textit{or} as a particle,
depending on the kind of experiment that is being performed, has generated
considerable confusion, the most well known case being the Wheeler
delayed-choice experiment \textrm{[8]}.\ In this experiment, single photons
impinge, successively, on a Mach-Zehnder (M-Z) interferometer.\textrm{\ }If
the second beam-splitter (BS) of the interferometer is in place
(\textquotedblleft closed\textquotedblright\ interferometer), interference can
be observed; for instance, we can have the photons always emerging on the same
output port; if it is removed (\textquotedblleft open\textquotedblright%
\ interferometer), the photons will always be found following one arm of the
interferometer or the other, never both at the same time \textrm{[9]}.
\textit{Apparently}, the photon behaves as a wave in the first case and as a
particle in the second. However, we are dealing with a quantum entity, not a
classical one, and the attempt to directly transpose our macroscopic
experience into the microworld may be misleading. Surely, a photon can
disclose wave-like and particle-like behaviors, which is not the same as being
able to \textquotedblleft adapt\textquotedblright\textit{\ }its behavior
according to the experimental\textit{\ }setup.\ The idea behind the Wheeler
delayed-choice experiment (although not necessarily explicitly assumed) is
that, by deciding to remove, or not, the second BS after the passage of the
photon through the first BS, we can try to \textquotedblleft
cheat\textquotedblright\ the photon, so to speak, which will not
\textquotedblleft know\textquotedblright\ if it is impinging on an open\ or a
closed\ M-Z interferometer. For instance, when it impinges on the first BS, we
keep the second BS removed, but, while it is still inside the interferometer,
we put the second BS in place. It is implicit conjectured that the photon
might \textquotedblleft decide\textquotedblright\ to behave as a particle or
as a wave when it impinges on the first BS (somehow it would \textquotedblleft
know\textquotedblright\ whether the second BS is in place or not)
\textrm{[10]}. Naturally, this appears to be a preposterous idea, and goes
against the essence of quantum mechanical formalism (QMF).\textrm{\ }According
to QMF, the impinging photon is represented by a ket that is split at the
first BS, and the way it is split does not depend on the second BS being in
place or not, that is, on the M-Z interferometer being open or closed. As has
been correctly pointed out recently \textrm{[11]}, delayed-choice experiments
do not imply retrocausality.\ In fact, what is essentially new -- and this has
to do with the nonlocal features of the quantum world -- is that whenever the
photon is detected on one of the arms of the interferometer the ket following
the other arm is nullified; on the other hand, if the photon is not detected
on one of the arms, it will necessarily be detected on the other (negative or
null-result experiment \textrm{[12]}). Situations in which a photon seems to
behave neither as a particle nor as a wave, making evident the impossibility
of tracing back the photon path have also been discussed \textrm{[13]}. We can
also consider a M-Z interferometer in which the mirrors are replaced by BSs,
which we can designate as an \textquotedblleft open-closed\textquotedblright%
\ interferometer. In this case, how is the photon supposed to behave? Will it
plunge into a state of internal confusion, incapable to decide what to do? Or,
will it take its decision (to behave as a particle or as a wave) before
impinging on the first BS? Undoubtedly, if we insist in following this line of
thought, we will end up adopting an animist vision of reality.

It is important to stress that ontic interpretations of the pilot-wave kind
\textrm{[14] }cannot a priori be discarded. In this case, particles would have
well-defined trajectories, being guided by waves. But, also here, we cannot
cheat this wave \textit{and }particle entity by doing a delayed-choice
experiment. In conclusion, delayed-choice experiments \textrm{[15] }may justly
be considered impressive technical achievements, but are of little use to
unveil the mysteries of the quantum world or to challenge realism
\textrm{[16]}.


\begin{thebibliography}{99}
\bibitem {10}Actually, according to Wheeler (ref. [8]): "No elementary
phenomenon is a phenomenon until it is a registered (observed) phenomenon";
and, \textquotedblleft The past has no existence except as it is recorded in
the present. By deciding what questions our quantum registering equipment
shall put in the present we have an undeniable choice in what we have the
right to say about the past\textquotedblright; and still, \textquotedblleft
Useful as it is under everyday circumstances to say that the world exists `out
there'\ independent of us, that view can no longer be
upheld.\textquotedblright\ Therefore, it is far from clear, if we assume this
anthropocentric attitude, what is the purpose of the delayed-choice
experiment. What difference does it make to put or to remove the second BS
after the photon has already passed the first BS? What is to be expected from
this? What does it prove (the confirmation of the quantum mechanical
predictions) with respect to the dispute between the ontological (realist) and
the epistemological (antirealist) points of view?
\end{thebibliography}
\end{document}